\documentclass[a4paper, 10pt]{article}
\usepackage{authblk}
\usepackage[font={small,it}]{caption}
\usepackage{lingmacros}
\usepackage{tree-dvips}
\usepackage{amsmath} 
\usepackage{multicol}
\usepackage{xcolor}
\usepackage{graphicx}

\title{X-ray Phase Measurements by Time-Energy Correlated Photon Pairs}
\author[1,2,3]{Y. Klein}
\author[1,3]{E. Strizhevsky}
\author[1,3]{H. Aknin}
\author[1]{M. Deutsch}
\author[4]{E. Cohen}
\author[1]{A. Pe'er}
\author[3]{K. Tamasaku}
\author[5]{T. Schulli}
\author[2,6]{E. Karimi}
\author[1,3]{S. Shwartz}

\affil[1]{Physics Department and Institute of Nanotechnology and advanced Materials, Bar Ilan University, Ramat Gan 52900, Israel}
\affil[2]{Nexus for Quantum Technologies, University of Ottawa, Ottawa, Ontario K1N 6N5, Canada}
\affil[3]{RIKEN SPring-8 Center, 1-1-1 Kouto, Sayo-cho, Sayo-gun, Hyogo 679-5148, Japan}
\affil[4]{Faculty of Engineering and Institute of Nanotechnology and advanced Materials, Bar Ilan University, Ramat Gan 52900, Israel}
\affil[5]{ESRF — The European Synchrotron, Grenoble 38043, France}
\affil[6]{Institute for Quantum Studies, Chapman University, Orange, California 92866, USA}

\date{}
\begin{document}
\maketitle

\begin{abstract}
\textbf{The invention of X-ray interferometers has led to advanced phase-sensing devices that are invaluable in various applications. These include the precise measurement of universal constants, e.g. the Avogadro number \cite{Becker2007ConsiderationsUnits,Becker2003TracingCrystal}, of lattice parameters of perfect crystals \cite{Basile1994MeasurementSpacing,Ferroglio2008SiInterferometry}, and phase-contrast imaging, which resolves details that standard absorption imaging cannot capture \cite{Davis1995Phase-contrastX-rays,Pfeiffer2006PhaseSources,Momose1996Phase-contrastTissues,Miao2016AImaging}. However, the sensitivity and robustness of conventional X-ray interferometers are constrained by factors, such as fabrication precision, beam quality, and, importantly, noise originating from external sources or the sample itself \cite{Yoneyama2023Crystal-BasedThermography}. In this work, we demonstrate a novel X-ray interferometric method of phase measurement with enhanced immunity to various types of noise, by extending, for the first time, the concept of the SU(1,1) interferometer \cite{Yurke1986SU2Interferometers} into the X-ray regime. We use a monolithic silicon perfect crystal device with two thin lamellae to generate correlated photon pairs via spontaneous parametric down-conversion (SPDC). Arrival time coincidence and sum-energy filtration allow a high-precision separation of the correlated photon pairs, which carry the phase information from orders-of-magnitude larger uncorrelated photonic noise. The novel SPDC-based interferometric method presented here is anticipated to exhibit enhanced immunity to vibrations as well as to mechanical and photonic noise, compared to conventional X-ray interferometers. Therefore, this SU(1,1) X-ray interferometer should pave the way to unprecedented precision in phase measurements, with transformative implications for a wide range of applications.}
\end{abstract} 

\section*{Introduction}
Interferometers are the most direct and sensitive instruments for phase measurements, playing a pivotal role in fundamental science and numerous applications. The demand for higher performance has driven the development of advanced interferometers, sparking great interest across a broad range of disciplines. The conventional approach to phase sensing with interferometers relies on amplitude-splitting a beam into two, each of which follows a different spatial path, and a subsequent recombination of these beams. Constructive or destructive interference occurs depending on the relative phase between the two waves. Thus, the detected intensity provides information on the relative phase between the waves.  

This approach was adapted to X-rays by Bonse and Hart \cite{Bonse1965AnInterferometer}, who replaced standard beam splitters and mirrors, used in the optical regime, by reflecting lattice planes in devices cut from perfect silicon crystals. As shown in Fig. 1(a), such an interferometer utilizes three crystal lamellae. The first splits the beam into two (transmitted and Laue-reflected), and the second Laue-reflects and recombines them to create an interference pattern on the third lamella, employed to detect intensity variations in the interference pattern when a phase object is inserted into one of the interfering beams. This lamella employs its \AA-scale atomic structure periodicity to detect the intensity variation in the interference pattern on the \AA-scale, dictated, in turn, by the X-rays’ \AA-scale wavelength.

Although this design can provide, in principle, a very high sensitivity to phase differences, it is highly susceptible to vibration and fabrication imprecisions due to the short wavelength \cite{Yoneyama2023Crystal-BasedThermography,Diemoz2012AnalyticalTechniques}. Monolithic crystal X-ray interferometers provide higher immunity to vibrations and mechanical noise, but impose limits on imaged object size and fabrication tolerances \cite{Yoneyama2023Crystal-BasedThermography,Diemoz2012AnalyticalTechniques}. Perfect crystal X-ray interferometers also mandate the use of highly monochromatic input beams that are extremely sensitive to various types of scattering and stray radiation, which can significantly degrade the quality of the information they provide. 
While other types of X-ray interferometers based on diffraction and propagation \cite{Momose1996Phase-contrastTissues,Miao2016AImaging,Modregger2011SensitivityInterferometry,Pfeiffer2008Hard-X-rayInterferometer} have demonstrated advantages in certain aspects, their phase sensitivity is limited compared to crystal-based X-ray interferometers \cite{Yoneyama2023Crystal-BasedThermography,Diemoz2012AnalyticalTechniques}.

Here, we demonstrate an alternative approach to X-ray interferometry based on coincidence measurements and the SU(1,1) interferometry approach \cite{Yurke1986SU2Interferometers, Jing2011RealizationAmplifiers}. This is the first implementation of coincidence measurements of highly correlated pairs for X-ray phase sensing, significantly improving the signal-to-noise ratio (SNR) of phase measurements. The SU(1,1) interferometer, previously implemented in optical experiments \cite{Yurke1986SU2Interferometers,Jing2011RealizationAmplifiers,Chekhova2016NonlinearOptics,Shaked2018LiftingAmplification,Michael2021AugmentingAsymmetry,Manceau2017DetectionInterferometer,Frascella2021OvercomingSensing,Meir2023UltrafastInterferometer}, but never with X-rays, offers several significant advantages over other types of interferometers. These advantages include robustness against vibrations \cite{Chekhova2016NonlinearOptics}, ability to use parametric gain to improve SNR \cite{Manceau2017DetectionInterferometer}, reduced sample dose \cite{Frascella2021OvercomingSensing}, and capability to overcome system losses and inefficient detection \cite{Frascella2021OvercomingSensing,Michael2021AugmentingAsymmetry,Manceau2019Indefinite-MeanNoise,Manceau2017ImprovingUnbalancing}.
Another advantage is its large bandwidth and acceptance angle, which could enable applications in phase-contrast imaging of nanoscale objects and interferometric measurement of ultrafast phenomena.

The operating principle of the SU(1,1) interferometer \cite{Yurke1986SU2Interferometers,Jing2011RealizationAmplifiers,Chekhova2016NonlinearOptics}, which relies on the phase dependence of nonlinear optical mixers and amplifiers, is illustrated in Fig. 1(b). When two nonlinear media are arranged sequentially to generate photon pairs e.g. through spontaneous parametric down-conversion (SPDC), the phase difference becomes measurable through the detection of the photons emerging from the second crystal. Specifically, the flow of energy in the second crystal - from the pump to SPDC or \textit{vice versa} depends on the relative phase between the SPDC photon-pairs and the pump ($\Delta\phi=\phi_s+\phi_i-\phi_p$). Quantum mechanically, the process is a bi-photon interference mechanism, where photons can be generated or annihilated only in pairs, which leads to nonclassical squeezing and is key to the utilization of SU(1,1) interference in quantum applications of sensing, communication, computing, etc. \cite{Michael2019Squeezing-enhancedSpectroscopy,Eldan2023MultiplexedBandwidth}. Introducing an object between the two crystals causes significant phase accumulation differences between the pump, signal, and idler beams due to dispersion, resulting, in turn, in a change in the photon count rate of the signal and idler. Varying this phase shift by varying, e.g., the object's thickness, results in intensity variations in the counters akin to interference patterns. The count rate variation enables, therefore, an accurate determination of the phase shift variation.
\begin{figure}
    \centering
    \includegraphics[width=1\linewidth]{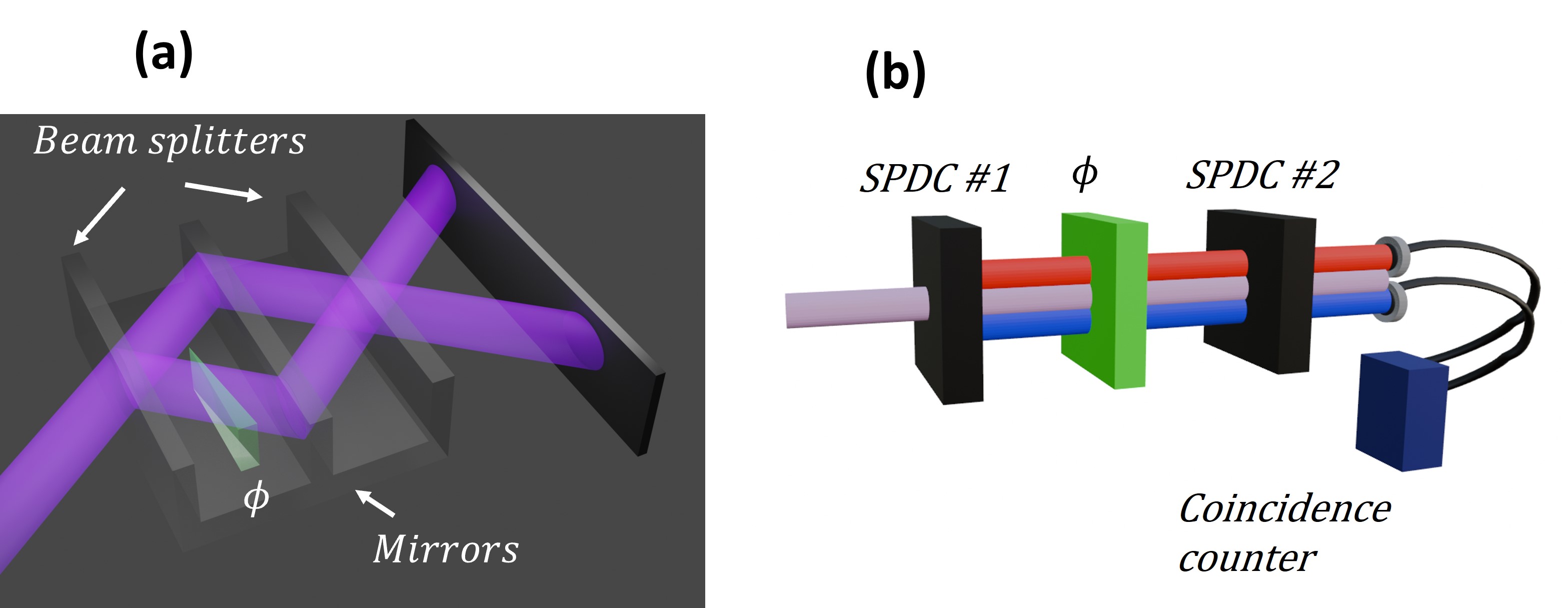}
    \caption{Comparison of Mach-Zehnder and SU(1,1) Interferometers. (a) X-ray Mach-Zehnder interferometer: Laue diffraction in the leftmost lamella of a monolithic crystal interferometer generates two beams that are Laue-reflected by the second lamella to interfere at the position of the third lamella, causing the resultant intensities to vary depending on the relative phase difference between the two waves. (b) SU(1,1) interferometer: Nonlinear crystals replace the beam splitters. These generate from the pump beam (purple) by parametric down-conversion two photons: idler (blue) and signal (red) detected by two detectors with coincidence discrimination. The intensity at the output (detectors) depends on the interference between the modes. A phase object placed between the two crystals introduces different phase shifts due to wavelength dispersion, yielding a corresponding change in the counted intensity. This allows for the detection of relative phase shifts by intensity measurements.}
    \label{fig:Basic_principle}
\end{figure}

\section*{Experimental setup and results}
The design is based on the key principles demonstrated in the SU(1,1) interferometers at optical wavelengths, with necessary modifications for X-rays \cite{Sofer2019QuantumDetection, Schori2018GhostPhotons,Strizhevsky2021EfficientSplitter} as depicted in Fig. 2(a) and described below. The two major differences between X-ray and conventional SPDCs are the nonlinear mechanism and the phase matching scheme. Here, the nonlinearity is based on a plasma-like nonlinearity \cite{Eisenberger1971X-RayConversion,Shwartz2012X-rayRegime,Shwartz2011PolarizationEnergies}, and the phase matching is based on the atomic-scale periodicity of crystals \cite{Freund1969ParametricRays} as shown in Fig. 2b. Both the nonlinearity and the phase matching scheme impose a geometry where the signal and idler beams propagate at angles nearly twice the Bragg angle with respect to the pump beam \cite{Freund1969ParametricRays}. To mitigate the absorption effect and facilitate small angular shifts between the photons for optimizing the beams’ overlap, we used a high energy, 35 keV, for the pump. Since ensuring beam overlap is crucial for SU(1,1) interferometers, we employ a monolithic silicon crystal \cite{Deutsch1985ElectronicSilicon, Deutsch1985AMethod} with two lamellae spaced 5 mm apart. As the Laue-reflecting planes in both lamellae are perfectly aligned in a monolithic crystal device, the need for separate relative alignments of two crystals is eliminated. A monolithic crystal device is also a perfect solution to the challenges associated with beam overlap. To satisfy phase matching, the interferometer was tuned to an angle of 55.85 mrad, deviating from the Bragg angle by 0.15 mrad. The phase objects inserted between the device’s lamellae were silicon membrane combinations of varying thicknesses ranging from 2 to 28 microns (see Supplementary Information for more details about the interferometer and phase objects).
\begin{figure}
    \centering
    \includegraphics[width=1\linewidth]{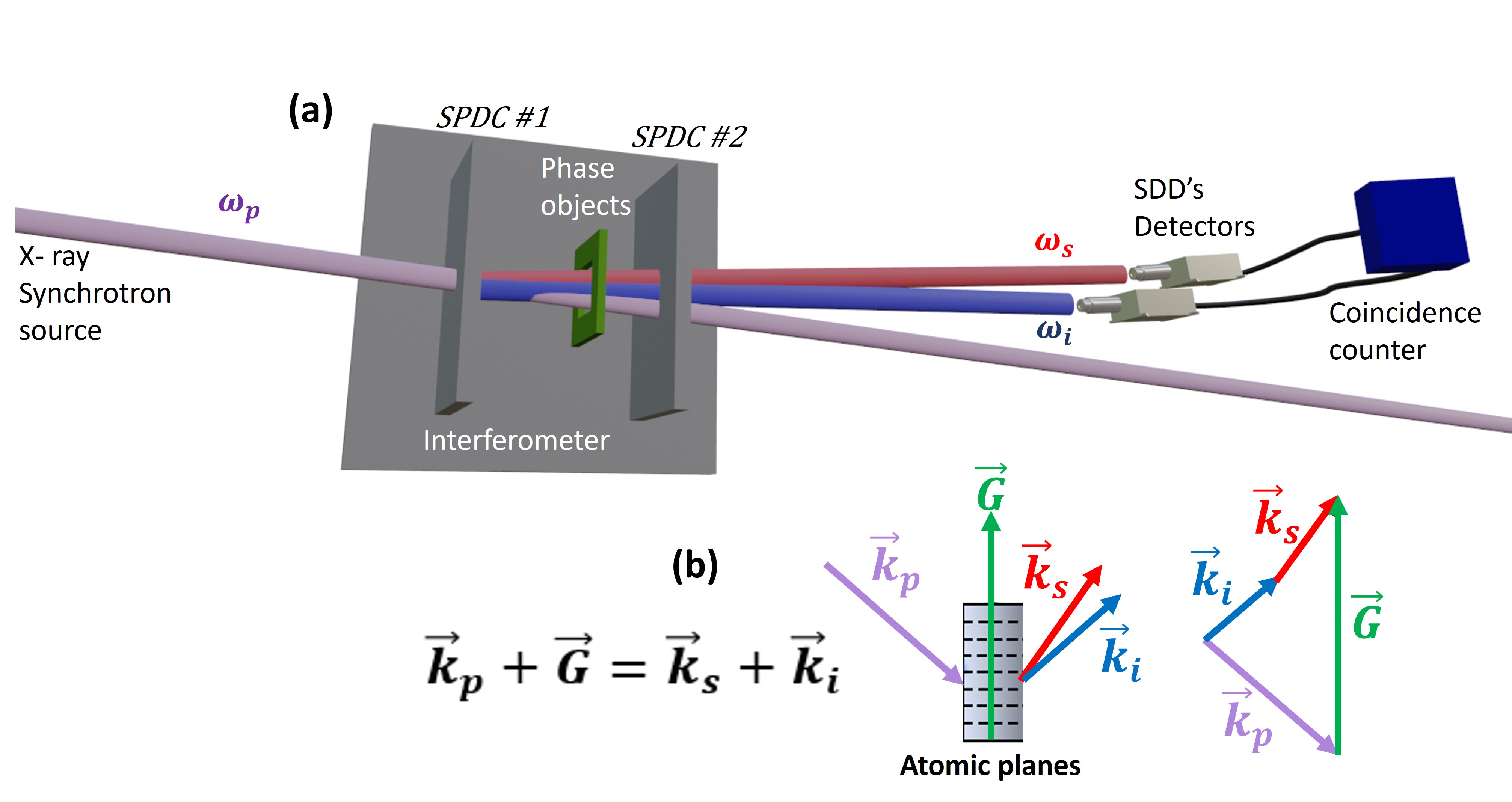}
    \caption{(a) Experimental scheme of the X-ray SU(1,1) Interferometer. A 35 keV beam enters the interferometer, constructed from a monolithic silicon crystal with two lamellae. The coincidence count rate at the detectors varies depending on the phase shift caused by a phase object inserted between the lamellae. (b) X-ray SPDC phase matching diagram where $\Vec{k}_p$, $\Vec{k}_s$ and $\Vec{k}_i$ are, respectively, the wave vectors of the pump, signal and idler, and $\Vec{G}$ is the lattice vector.}
    \label{fig:Experimental_setup}
\end{figure}

One of the major advantages of quantum sensing is its capability to utilize multiple correlations across various degrees of freedom of the emerging photons to enhance the SNR of the measured signal \cite{Szuniewicz2023Noise-resistantCorrelation,Thekkadath2023IntensityLight,Sofer2019QuantumDetection}. In this work, we primarily focus on time, energy, and propagation angle. To demonstrate this, we recorded both the time and the energy of each detected photon (By resolution that are far exceed the uncertainly limit), then post-selected photon pairs (detected at phase-matching angles) complying with energy conservation and analyzed the temporal distribution of their time difference. This suggested that coincidence measurements could improve the SNR by a factor inversely proportional to the biphoton correlation time, which could be as small as 100 zeptoseconds for X-rays. However, practical enhancement of the SNR is limited by the response time of the detectors, which is several orders of magnitude larger than the envelope of the biphoton states. Despite this inherent limitation, leveraging the temporal distribution can effectively contribute to the reduction of background noise, as we now show.  

Figure 3 demonstrates the impact of energy and temporal filtering. The average raw spectrum measured in one detector is shown in Fig. 3(a). Energy filtering, utilizing the photon energy resolving capabilities of our detectors to select photons in the range of 14 to 21 keV, is shown in blue. Note that this is not the full range of the SPDC, but the range that optimizes the SNR for our specific setup. Then, we used a fast digitizer to record the signals from both detectors within a 1000 ns window. The time difference histogram is plotted in Fig. 3(b), showing the zero-time difference bin to be higher than the others, but it is still difficult to distinguish the photon pairs from the background. 
\begin{figure}
    \centering
    \includegraphics[width=1\linewidth]{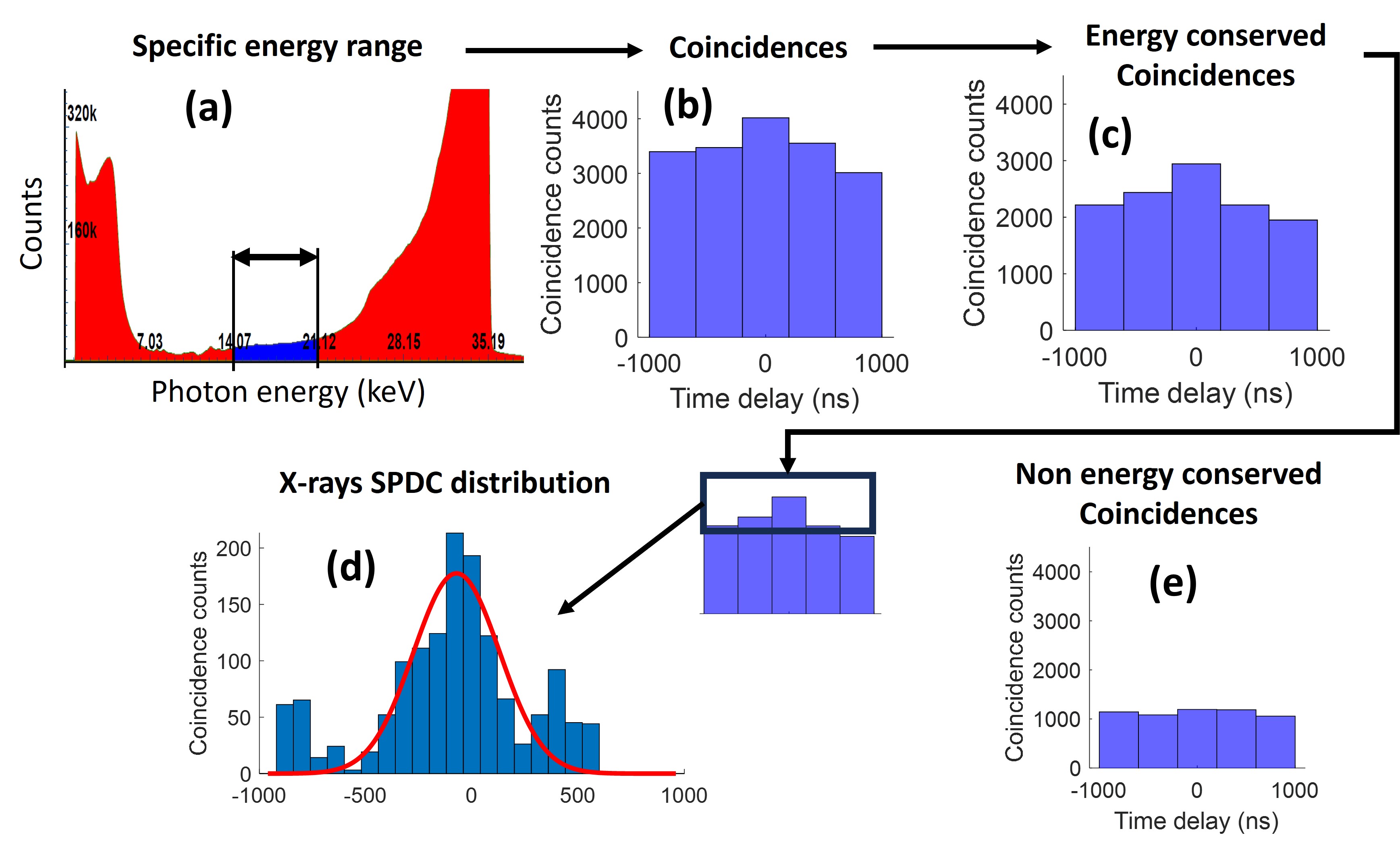}
    \caption{The process of noise filtering (details are discussed in the main text). (a) Full spectrum as measured by a single detector. The blue shadow indicates the energy range selected for SNR optimization. (b)-(e) Shows time difference histograms: (b) with time coincidence, but without energy conservation discrimination; (c) same as (b) but with energy conservation discrimination; (d) the data in (b) after the subtraction of the constant baseline spanned by the two outermost bins in (b). This baseline is generated by accidental coincidences, which are independent of the time difference. The red curve is a Gaussian fit yielding the actual temporal resolution of our system: 200 ns HWHM. (e) Counts that do not conserve energy exhibit a nearly uniform distribution, confirming random arrival times of background photon pairs.}
    \label{fig:Counts_histograms}
\end{figure}

We know that the sum of the energies of the generated photons is equal to the photon energy of the pump (corresponding to energy conservation). Applying this rule leads to the results shown in Fig. 3(c), which presents a Gaussian distribution on top of a nearly constant background. Next, we subtract the average of the two extremal bins that represent the uniform background distribution to better observe the Gaussian distribution of the SPDC photons (Fig. 3(d)). For reference, we show the counts that do not satisfy energy conservation in Fig. 3(e). A Gaussian fit to the data in Fig. 3(d) yields a \(\sim\)200 ns response time for our detectors. This procedure allows us to observe small phase variations even in the presence of significantly stronger noise levels.

Following the procedure described above, we conducted a series of coincidence measurements with the silicon membrane phase objects, of thicknesses varying from 0 to 28 \(\mu m\), inserted into the beam path. The key results of this study are displayed in Fig. 4 as symbols, accompanied by corresponding error bars. The red line in Fig. 4(a) represents the theoretically calculated curve, which will be discussed in the next section. To further validate the results in Fig. 4(a), we present Fig. 4(b), which shows the counts within the same time window as the photon pairs, but which do not satisfy energy conservation, demonstrating that the observed results are not coincidental. Clearly, no phase-shift dependence is observed in this case. 

Remarkably, the results in the insets indicate that the background noise varied between different membranes and was inconsistent even for repeated measurements on the same 24 microns thickness membrane. This variation arose from our use of Kapton tape to prevent membrane adhesion, where we deliberately varied the amount of Kapton in each measurement to introduce controlled variability, demonstrating the robustness of our measurement technique. These findings suggest that our method is resilient to scattering from the sample, even when the scattering is significant. Additionally, the tape introduced random inhomogeneous phases that fluctuated during the measurement. Despite these fluctuations, our phase measurements remained unaffected, further highlighting the robustness of the technique.
\begin{figure}
    \centering
    \includegraphics[width=1\linewidth]{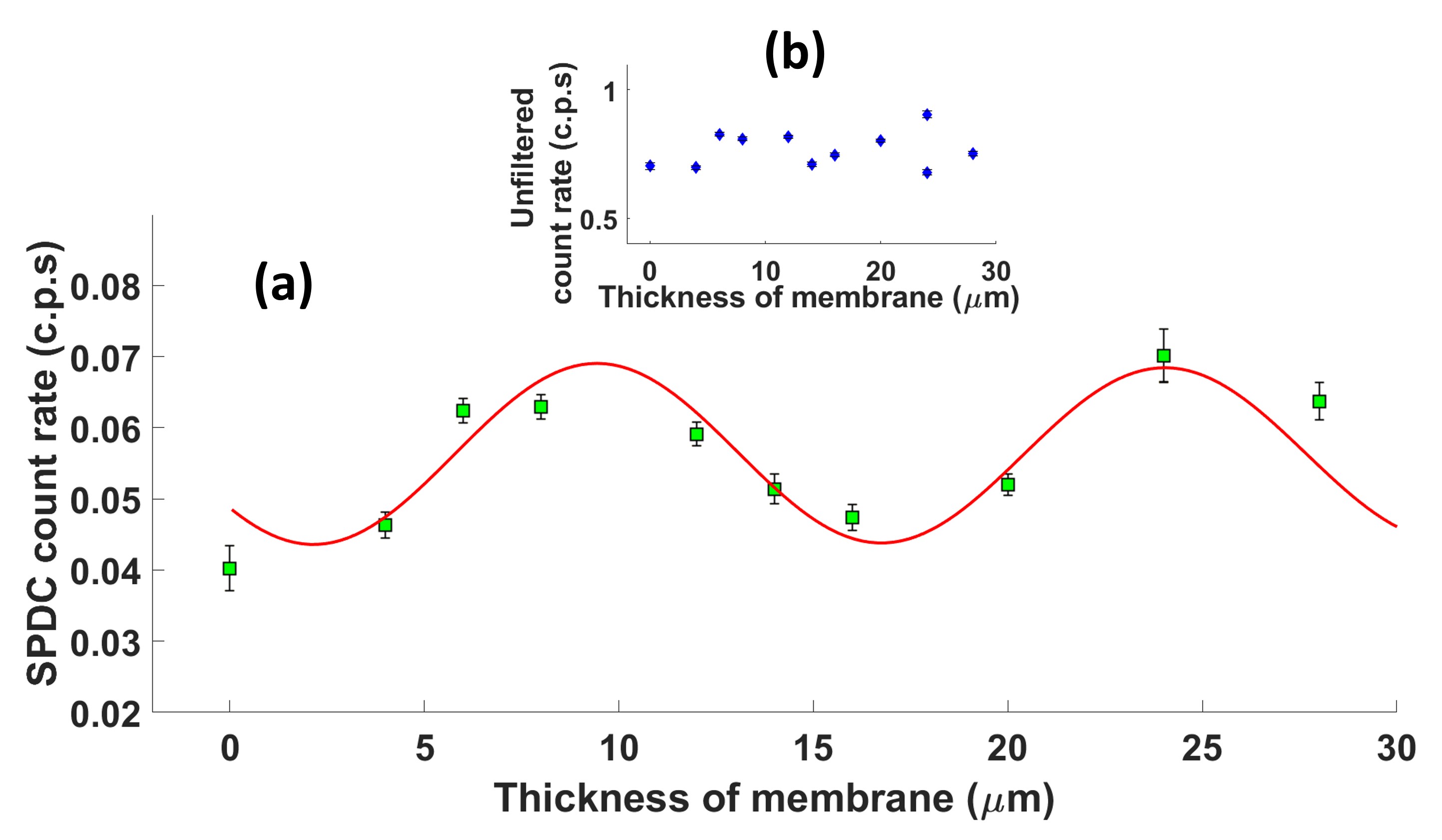}
    \caption{Coincidence count rate at the output of the second lamella dependence on the thickness of the membranes. (a) Measured SPDC count rate after using temporal and energy filtering (green squares) and theoretical curve (red line). The measurement time for each point was \(\sim\)6 hours. (b) Count rate of photon pairs within the coincidence time window, but without energy conservation filtering.}
    \label{fig:Results}
\end{figure}

\section*{Comparison with theory}
To calculate the coincidence count rate, we used the Glauber correlation function:
\begin{equation}\label{eq:1}
G^{(2)}(\tau,\mathbf{\mu})=\langle a_{si}^\dag (t_2,\mathbf{r_2}) a_{id}^\dag(t_1,\mathbf{r_1})a_{id} (t_1,\mathbf{r_1}) a_{si}(t_2,\mathbf{r_2}) \rangle
\end{equation}
at the output of the second lamella. Here, \(a_{si}\) and \(a_{id}\) are signal and idler annihilation operators, respectively, \(\mathbf{\mu}=\mathbf{r_2}-\mathbf{r_1}\) and \(\tau=t_2-t_1\).  The count rates were obtained by numerically integrating Eq. \eqref{eq:1} over the spectral and angular bandwidths. The operators are calculated in the frequency domain by considering the propagation through the three elements (the two lamellae and the phase object), including the loss and Langevin operators. This approach ensures the preservation of the commutators. The X-ray SPDC process was computed by solving the coupled wave equations for the signal and idler \cite{Shwartz2012X-rayRegime} in each of the lamellae:
\begin{equation}
\begin{split}
 & \frac{\partial {a}_{si}}{\partial z}+\frac{\alpha_{si}}{cos(\theta_{si})}a_{si}=\kappa a_{id}^\dag e^{i\Delta k z}+\sqrt{\frac{2\alpha_{si}}{cos(\alpha_{si})}}f_{si}
 \\ & \frac{\partial {a}^\dag_{id}}{\partial z}+\frac{\alpha_{id}}{cos(\theta_{id})}a^\dag_{si}=-\kappa^* a_{si} e^{i\Delta k z}+\sqrt{\frac{2\alpha_{id}}{cos(\alpha_{id})}}f^\dag_{id}
\end{split}.
\end{equation}
Here, \(\theta_{si}\) and \(\theta_{id}\) are the signal and idler angles; \(\alpha_{si}\) and \(\alpha_{id}\) are the absorption coefficients at the signal and idler frequencies, respectively, \(f_{si}\) and \(f_{id}\) are the Langevin noise operators, and \(\kappa\) is the nonlinear coupling coefficient. The phase of the object placed between the lamellae was represented by \(e^{i\Delta\phi}\) where the phase shift \(\Delta\phi\) is given by \cite{Chekhova2016NonlinearOptics}: 
\begin{eqnarray}
\Delta\phi&=&\phi_{si}+\phi_{id}-\phi_p\cr
    &=&2\pi d\left(\frac{n_{si}}{\lambda_{si}}+\frac{n_{id}}{\lambda_{id}}-\frac{n_p}{\lambda_p}\right),
\end{eqnarray}
and \(\lambda_p,\lambda_{si},\lambda_{id},n_p,n_{si},n_{id}\) and \(d\) are, respectively, the wavelengths, the silicon membrane’s indices of refraction for  the pump, signal, and idler, and the thickness of the membranes. For further mathematical details, see the Supplementary Information.

The experimental results shown in Fig. 4 are in good agreement with the calculated curve. This agreement required, three adjustments of the theoretical curve: a vertical shift, a count rate scaling, and a horizontal shift. These correspond, respectively, to a non-perfect overlap between the beams, uncertainties in the pump flux, and a possible  phase shift between the signal and idler and the pump even when no phase object is present in the space between the lamellae. The scaling is consistent with a pump flux of around \(10^{12}\) photons per second, in good agreement with the flux of the synchrotron beam. The comparison between the experiment results and the theory indicates that the phase shift in the absence of the membranes is \(\sim\pi/3\) corresponding to the gap between the two lamellae. The required vertical shift was 0.043 counts per second. The simulated visibility is 0.93, assuming an ideal beams overlap, whereas in our experiment, we observed a visibility of approximately 0.27. This is in consistent with approximately a \(30\%\) overlap of the beams in the experiment. The imperfect overlap is due to phase-matching requirements, introducing an angle of \(\sim\)0.1 radians between the pump beam and the generated photons, with the pump beam size of \(\sim\)0.9 mm and a lamellae distance of 5 mm. In spite of this overlap deficiency, which reduces the intensity variation visibility, the efficient background noise elimination by coincidence and photon energy discrimination allows precise measurements of very small phase variations with the present method.

\section*{Discussion and Conclusion}

We have achieved the first realization of X-ray quantum nonlinear interferometry, using SPDC of X-ray pump photons to generate correlated idler and signal photon pairs. In our setup, dispersion-induced phase shifts accumulated by the three modes as they traverse a phase object are converted into intensity variations at the interferometer’s output, enabling precise determination of phase shifts. We have leveraged the inherent pairing of photons in the SPDC process to demonstrate that coincidence and energy conservation filtering of these correlated photons allows for highly sensitive phase measurements, even in environments with significant noise. This powerful noise discrimination, made possible by our ability to identify photon pairs within a very noisy environment despite the very weak gain of the SPDC sources, is a uniquely quantum phenomenon with no classical analog, as demonstrated in several experiments in the optical range\cite{Vered2015Classical-to-quantumMixing,Nechushtan2021OptimalInterference}.

Although our focus was on time and energy correlations, other degrees of freedom, such as position or momentum (k-vector) correlations, could further enhance the SNR. The use of fast pixelated detectors, such as Medipix3 \cite{Ballabriga2011Medipix3:Performance,Sriskaran2024High-rateCapabilities,Goodrich2023ImagingProcess}, could facilitate this advancement. The main practical limitations of the scheme presented here are the time and energy resolutions of the coincidence measurement system. Notably, a recent experiment has demonstrated improved performance with enhanced electronics \cite{Hartley2024ConfirmingCorrelation}, suggesting that the SNR could be improved by approximately a factor of 10 compared to our current results. Further improvements will likely require advancements in detector technology. Finally, we note that phase measurements using intensity correlations with classical synchrotron radiation at X-ray wavelengths have been demonstrated \cite{Tamasaku2002X-RayCorrelation}. However, quantum correlations of the type used here are anticipated to provide even stronger results. 

 Our scheme enables the measurement of phase information even within optically opaque materials, expanding the scope and applicability of SU(1,1) interferometry to previously unexplored domains. For X-rays, this type of interferometer is very appealing since it eliminates the need for an analyzer crystal to detect sub-wavelength spatial shifts in the interference pattern. Furthermore, since the phase variation in an SU(1,1) interferometer is imprinted in the phase matching variation rather than in spatial variation, the interferometer is expected to be more stable against mechanical vibrations \cite{Chekhova2016NonlinearOptics}. By eliminating the need for stringent stability requirements hampering the performance of conventional X-ray interferometers, our method unleashes the potential of X-ray quantum interferometry, promising unparalleled sensitivity in quantum metrology. Consequently, we foresee our work laying foundations for the implementation of X-ray crystal interferometry with separate crystals, overcoming a significant limitation in this field. Extending our work to phase contrast imaging with a pixelated detector is straightforward and holds promise for highly sensitive imaging methods. This is because phase contrast often surpasses transmission contrast in many samples. The potential of this approach has already been demonstrated in the visible range \cite{Szuniewicz2023Noise-resistantCorrelation,Thekkadath2023IntensityLight}.

\section*{Methods}
Introductory measurements were carried out at the Nano/Micro Diffraction Imaging beamline (ID1) at ESRF \cite{Leake2017CoherentSynchrotron}. The results shown in Figs. 3 and 4 were obtained at the RIKEN SR physics beamline (BL19LXU) of the SPring-8 facility \cite{Yabashi2001Design1}. We employed a Si (111) monochromator to narrow the bandwidth of the input beam to approximately 1 eV. The spot size of the pump beam was defined by slits of 0.9 x 0.3 mm$^2$ (horizontal $\times$ vertical). To avoid detector count-rate saturation, the input flux was reduced by an aluminium absorber. The output photons, of energies centered at 17.5 keV, were measured by two energy-resolving photon-counting Amptek XR-100SDD Silicon Drift Detectors (SDDs). The detectors were connected to a PicoScope 6407 high-speed digitizer for data analysis. The coincidence time window of the detectors was 1000 ns. For the energy conservation discrimination the tolerance on the photon energies was $\pm3$ keV.

\section*{Funding} This work is supported by the Israel Science Foundation (ISF) (847/21); Y.K. and E.K. acknowledge the support of Canada Research Chair and Quantum Sensors Challenge Program at the National Research Council of Canada.
\bibliography{references}
\bibliographystyle{naturemag}
\end{document}


\maketitle

\section{Model for X-ray SU(1,1) interferometer}
Here, we provide the theoretical background for our simulation of the SU(1,1) interferometer results shown in  red line in Fig. 4 of the main text. We first derive Eqs.(2) of the main text, which establish the mathematical framework for spontaneous parametric down-conversion (SPDC) of X-rays. Next, we will proceed to discuss the SU(1,1) interferometer model, and to describe the methodology employed in our simulation.

\subsection{Wave equation}
We are starting with the scalar wave equation for the electric field, assuming no external charges or currents:
\begin{equation}
\label{wave equation}
\nabla^2 E = \mu_0 \epsilon_0\frac{\partial^2E}{\partial t^2}+\mu_0\frac{\partial^2P}{\partial t^2}+\mu_0\frac{\partial J}{\partial t}.
\end{equation}
Here, \(\epsilon_0\) and \(\mu_0\) are the vacuum permittivity and permeability, respectively. E, P, and J denote the electric field, the polarization, and the displacement current, respectively. We Fourier transform Eq. \eqref{wave equation} as: 
\begin{equation}
E(x,y,z,t)= \int_{-\infty}^{\infty} \int_{-\infty}^{\infty}\widetilde E(z,k_x,k_y,\omega)e^{-i(\omega t-k_xx-k_yy)}dk_xdk_yd\omega,
\end{equation}
yields the wave equation in the frequency domain:
\begin{equation}
\label{wave equation F}
\frac{\partial^2\widetilde E}{\partial z^2}-k_x^2\widetilde E -k_y^2\widetilde E=-\mu_0\epsilon_0 \omega^2\widetilde E-\mu_0\omega^2\widetilde P-i\mu_0\omega\widetilde J.
\end{equation}
Here, \(k_j\) is the wave vector component in the Cartesian direction j, and \(\omega\) is the angular frequency. Using \(k=\omega n/{c}\) (\(n\) and \(c\) are the real index of refraction and the speed of light, respectively) and the polarization \(P(\omega)=\epsilon_0[n(\omega)^2-1]E(\omega)\), Eq. \eqref{wave equation F} becomes: 
\begin{equation}
\label{wave equation F2}
\frac{\partial^2\widetilde E}{\partial z^2}-k_x^2\widetilde E -k_y^2\widetilde E=-k^2\widetilde E-i\mu_0\omega\widetilde J.
\end{equation}
Using \(k^2=k_x^2+k_y^2+k_z^2\) yields the wave equation: 
\begin{equation}
\label{Final wave equation}
\frac{\partial^2\widetilde E}{\partial z^2}+k_z^2\widetilde E =-i\mu_0\omega\widetilde J.
\end{equation}

\subsection{Slowly Varying Envelope Approximation}
Since the nonlinear interaction is weak, the variation of the electric field along the z-axis is slow. We may then write: 
\begin{equation}
\label{SVEA assumption}
\widetilde{E} = \mathcal{E}(z, k_x, k_y, \omega)e^{ik_zz}, \quad \widetilde{J} = \mathcal{J}(z, k_x, k_y, \omega)e^{iq_zz},
\end{equation}
with the assumption that \(\frac{\partial^2 \mathcal{E}}{\partial z^2} \ll k_z \frac{\partial \mathcal{E}}{\partial z}\). Substituting Eq. \eqref{SVEA assumption} into Eq. \eqref{Final wave equation} and neglecting second-order terms, as per the Slowly Varying Envelope  approximation (SVEA), yields:
\begin{equation}
\label{SVEA equation1}
2k_z \frac{\partial \mathcal{E}}{\partial z} = -\mu_0 \omega \mathcal{J}.
\end{equation}
Using  the definition of the impedance \(\eta =\frac{\sqrt{\mu_0/\epsilon_0}}{n}\) and the angle of propagation \(\theta_z\) relative to the z-axis, so that \(\cos(\theta_z) = k_z/k\), Eq. \eqref{SVEA equation1} becomes: 
\begin{equation}
\label{SVEA equation}
\cos(\theta_z) \frac{\partial \mathcal{E}}{\partial z} = -\frac{\eta}{2} \mathcal{J}.
\end{equation}

\subsection{Plasma-like nonlinearity}
We use the model introduced by Eisenberger and McCall \cite{Eisenberger1971X-RayConversion,Shwartz2011PolarizationEnergies} for the nonlinear interaction between light and matter. Using this model, we assume that since the hard X-ray photon energy is much higher than
the binding energy of the material’s electrons, we can treat the matter like a cold, dense plasma. For instance, in our experiment, the lowest X-ray energy was around 14 keV, compared to the binding energy of silicon electrons, which is only about 1.6 keV.  In this model, the nonlinear current density—arising from three components: the Lorentz force, a convective term, and the electric-field-induced charge density variation—serves as the source of the nonlinearity. The nonlinear current density is calculated by solving the equation of motion for a single electron (Eq. S.9), along with the continuity equation (Eq. S.10) for nonlinear velocity and charge density.
\begin{equation}
\frac{\partial \vec v}{\partial t}(\vec v\cdot \nabla)\vec v=-\frac{\mathcal{Q}_e}{m}(\vec E+\vec v \times \vec B)
\end{equation}
\begin{equation}
\frac{\partial \rho}{\partial t}+\nabla \cdot (\rho \vec v)=\frac{\partial \rho}{\partial t}+\nabla\cdot \vec J=0.
\end{equation}
Here, \(\mathcal{Q}_e\) and m are, respectively, the electron's charge and mass, \(\rho(\vec r)\) is the electron density function, and \(\vec B\)  and \(\vec v\) are, respectively, the magnetic field and the electron velocity vectors.

These equations can be solved perturbatively to obtain the nonlinear current density at the signal frequency
\begin{equation}
\label{NL current vector}
\begin{split}
 \vec J_{si}^{NL}(\vec r,t) = & \frac{i\mathcal{Q}_e^2}{m^2}\left[-\frac{\rho(\vec r)}{\omega_{si} \omega_{id} \omega_p} \nabla (\vec E_p(\vec r,t) \cdot \vec E^*_{id}(\vec r,t))\right.\\ & -\frac{1}{\omega_{id} \omega_p^2}(\nabla \rho(\vec r)\cdot \vec E_p(\vec r,t)) \vec E^*_{id}(\vec r,t) \\ & \left.+\frac{1}{\omega_{id}^2 \omega_p}(\nabla\rho(\vec r) \cdot \vec E^*_{id}(\vec r,t)) \vec E_p(\vec r,t)\right].
\end{split}
\end{equation}
Since the matter here is a crystal with a periodic electron distribution, \(\rho(\vec r)\) can be written as \(\rho(\vec r)=\sum_{g} \rho_g \exp{(i \vec G_g \cdot \vec r)}\), where \( \vec G_g \) is the reciprocal lattice vector. In the experiment, we exploit this periodic structure for phase matching by selecting a particular reciprocal lattice vector by precisely tuning the angles of the crystal and the detectors to satisfy the condition \(\vec G +\vec {k_x}_p=\vec {k_x}_{si}+\vec {k_x}_{id}\). Applying this condition and using the SVEA Eq. \eqref{NL current vector} becomes
\begin{equation} \label{NL current scalar}
\begin{split}
  \widetilde {J}_{si}^{NL} = \frac{\mathcal{Q}_e^2}{m^2}\frac{\omega_s \rho_g \widetilde E_p \widetilde E_{id}^*}{\omega_{si}^2 \omega_{id}^2 \omega_p^2} &\left[-\omega_{id} \omega_p(\vec e_p\cdot \vec e_{id})(\vec G \cdot e_{si})\right.\\ & -\omega_{si} \omega_{id}(\vec e_{id} \cdot \vec e_{si})(\vec G \cdot e_{p}) \\ & \left.+\omega_{si} \omega_p(\vec e_p \cdot \vec e_{si})(\vec G \cdot e_{id})\right].
\end{split}
\end{equation}
 
Since the signal and idler are nearly colinear, the second term in Eq. \eqref{NL current scalar} is  the dominant term and using \ref{SVEA assumption} yields:
\begin{equation}
{\mathcal{J}}^{NL}_{si} = \omega_{si} \zeta {\mathcal{E}}_{p}  {\mathcal{E}}^*_{id} e^{i\Delta k_zz}
\end{equation}
where \(\zeta=[q_e^2\rho_g/(m^2\omega_{si} \omega_{id} \omega_p^2)] (\vec G \cdot e_{p})\) and \(\Delta k_z = {k_z}_p-{k_z}_{si}-{k_z}_{id}\). 
\color{black}
Following the same procedure for the pump and signal currents, we obtain: 
\begin{equation}\label{three coupled plasma NL}
\begin{split}
 & {\mathcal{J}}^{NL}_{si} = \omega_{si} \zeta {\mathcal{E}}_{p}  {\mathcal{E}}^*_{id} e^{i\Delta k_zz}
 \\ & {\mathcal{J}}^{NL}_{id} = \omega_{id} \zeta {\mathcal{E}}_{p}  {\mathcal{E}}^*_{si} e^{i\Delta k_zz}
 \\ & {\mathcal{J}}^{NL}_{p} = \omega_{si} \zeta {\mathcal{E}}_{id}  {\mathcal{E}}_{si} e^{i\Delta k_zz}
 \end{split}
\end{equation}
By substituting Eq. \eqref{three coupled plasma NL} into Eq. \eqref{SVEA equation}, assuming propagation in a lossy medium with absorption coefficient \(\alpha_j\) and applying the undepleted pump approximation (i.e. a constant \(\mathcal{E}_p\)) the  classical coupled equations are: 
\begin{equation} \label{two wave NL}
\begin{split}
 & cos(\theta_{si})\frac{\partial {\mathcal{E}}_{si}}{\partial z}+\alpha_{si}{\mathcal{E}}_{si}=-\frac{\eta}{2}\omega_{si} \zeta {\mathcal{E}}_{p}  {\mathcal{E}}^*_{id} e^{i\Delta k_zz}
 \\ & cos(\theta_{id})\frac{\partial {\mathcal{E}}_{id}}{\partial z}+\alpha_{id}{\mathcal{E}}_{id}=-\frac{\eta}{2}\omega_{id} \zeta {\mathcal{E}}_{p}  {\mathcal{E}}^*_{si} e^{i\Delta k_zz}
\end{split}
\end{equation}
These equations describe the propagation of the signal and idler fields in the presence of nonlinear coupling and absorption. 

Fig. S1 provides a description of vectors, axes, and angles. As one example we show \(\theta_p\) which is the angle between the direction of the pump k-vector and axis z. The same will be for \(\theta_{id}\) and \(\theta_{si}\).
\begin{figure}
    \centering    \includegraphics[width=0.75\linewidth]{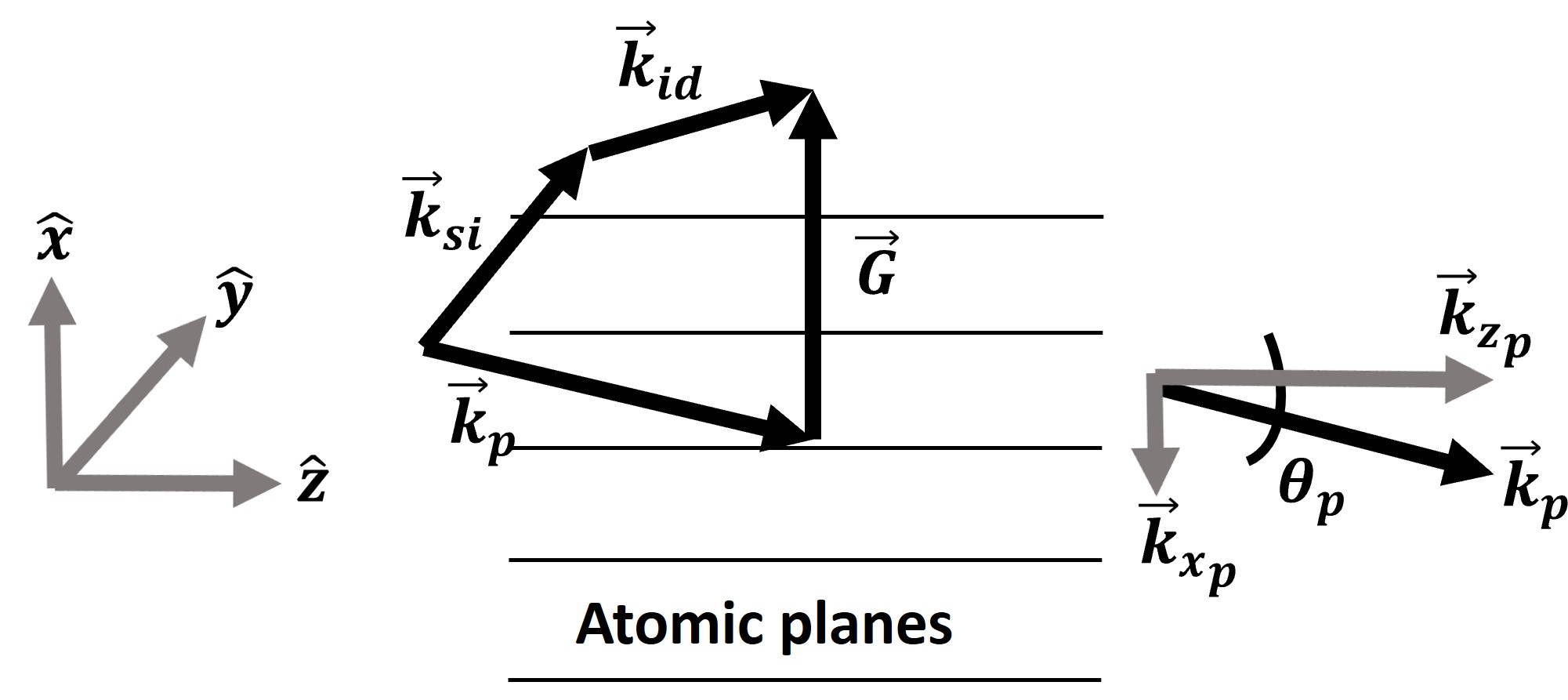}
    \caption{definition of vectors, axis and angles}
    \label{fig:SPDC_axis}
\end{figure}

\subsection{Quantization}
We employ the Glauber second-order correlation function to calculate the coincidence count rates. To do this, we quantize the envelopes of the signal and idler fields, ensuring that they are normalized so that the number operator corresponds to the photon flux at the z plane, so that the operators satisfy the relation \( {\mathcal{E}}_{j}=\sqrt{\omega_j\eta_j/cos\theta_j}a_j\). Working in the frequency and transverse k-vector domain, the commutators can be written as:
\begin{equation}
\begin{split}
[a_j(z_1, q_1, \omega_1),& a_k^\dag(z_2, q_2, \omega_2)]=\\&\frac{1}{(2\pi)^3}\delta_{j,k}\delta(z-z')\delta(q-q')\delta(\omega-\omega'),
\end{split}
\end{equation}
where \(q=\{k_x,k_y\}\). 
 
To preserve these commutators, we apply the Langevin approach \cite{Shwartz2012X-rayRegime} and introduce noise operators with the commutators
\begin{equation}
\begin{split}
[f_j(z_1, q_1, \omega_1),& f_k^\dag(z_2, q_2, \omega_2)]=\\&\frac{1}{(2\pi)^3}\delta_{j,k}\delta(z-z')\delta(q-q')\delta(\omega-\omega'),
\end{split}
\end{equation}
to the coupled equations \ref{two wave NL} to obtain:
\begin{equation} \label{MD}
\begin{split}
 & \frac{\partial {a}_{si}}{\partial z}+\frac{\alpha_{si}}{cos(\theta_{si})}a_{si}=\kappa a_{id}^\dag e^{i\Delta k z}+\sqrt{\frac{2\alpha_{si}}{cos(\alpha_{si})}}f_{si}
 \\ & \frac{\partial {a}^\dag_{id}}{\partial z}+\frac{\alpha_{id}}{cos(\theta_{id})}a^\dag_{si}=-\kappa^* a_{si} e^{i\Delta k z}+\sqrt{\frac{2\alpha_{id}}{cos(\alpha_{id})}}f^\dag_{id},
\end{split}
\end{equation}
with the coupling coefficient 
\(\kappa=(\zeta/2) [(\omega_{si}\omega_{id}\omega_p\eta_{si}\eta_{id}\eta_p)/(\cos{\theta_{si}}\cos{\theta_{id}}\cos{\theta_{p}})]^{1/2}\).  
These are Eqs. (2) in the main text and describe X-ray SPDC in a lossy medium \cite{Shwartz2012X-rayRegime,Balic2005GenerationWaveforms,Kolchin2007Electromagnetically-induced-transparency-basedGeneration,RaymondOoi2007ContinuousLaser,RaymondOoi2007CorrelationTheory,Du2008NarrowbandResonance}.

\subsection{Solution approach}

To solve the coupled equations \ref{MD}, we first transform them into equations with constant coefficients by introducing \(a_s=b_s \exp(i\Delta k_zz/2)\) and \(a_i^\dag=b_i^\dag \exp(-i\Delta k_zz/2)\). This yields:
\begin{equation}	
\begin{split}
& \frac{\partial b_{si}}{\partial z} +\left(\frac{\alpha_{si}}{\cos\theta_{si}}+\frac{i\Delta k_z z}{2}\right)b_{si} = \kappa b_{id}^\dag +\sqrt{\frac{2\alpha_{si}}{cos\theta_{si}}}f_{si}
\\ & \frac{\partial b_{id}^\dag}{\partial z} +\left(\frac{\alpha_{id}}{\cos\theta_{id}}-\frac{i\Delta k_z z}{2}\right)b_{id}^\dag = \kappa^* b_{si} +\sqrt{\frac{2\alpha_{id}}{cos\theta_{id}}}f_{id}^\dag.
\end{split}
\end{equation}

For a crystal of length L, we define a matrix M and an exponential \(\exp(-ML)\) as:
\begin{equation}
\begin{split}
& M=\begin{bmatrix}
{\alpha_{si}+i\Delta k_z /2} & \kappa \\
\kappa & {\alpha_{id}-i\Delta k_z /2}
\end{bmatrix}   
\\ &   \exp(-ML)=\begin{bmatrix}
A & B \\
C & D
\end{bmatrix}.
\end{split}
\end{equation}
By integrating over \(z\) we obtain
\begin{equation}
\begin{split}
\begin{bmatrix} {b_{si}}(L) \\ {b_{id}^\dag}(L)\end{bmatrix}=& \exp(-ML) 
\begin{bmatrix} {b_{si}(0)} \\ {b_{id}^\dag(0)}\end{bmatrix}
\\ & +\sqrt{2}\int_{0}^{L} \exp(-M(s-L)\begin{bmatrix} {\sqrt{\alpha_{si}}f_{si}(s)} \\ {\sqrt{\alpha_{id}}f_{id}^\dag(s)}\end{bmatrix} \,ds .
\end{split}
\end{equation}
Expanding the Matrix and keeping terms only up to the first order in \(\kappa\), we obtain
\begin{equation}
\label{Crystal operators solution}
\begin{split}
& {b_{si}}(L)= A{b_{si}(0)}+B{b_{id}^\dag(0)}
\\& \qquad +\int_{0}^{L} E(s)\sqrt{\alpha_{si}}f_{si}(s)\,ds+\int_{0}^{L} F(s)\sqrt{\alpha_{id}}f_{id}^\dag(s)\,ds
\\& {b_{id}}^\dag(L)= C{b_{si}(0)}+D{b_{id}^\dag(0)}
\\& \qquad +\int_{0}^{L} G(s)\sqrt{\alpha_{si}}f_{si}(s)\,ds+\int_{0}^{L} H(s)\sqrt{\alpha_{id}}f_{id}^\dag(s)\,ds
\end{split},
\end{equation}
where
\begin{equation}
\exp[-M(s-L)]=\begin{bmatrix}
E & F \\
G & H
\end{bmatrix}.
\end{equation}

\subsection{Count rates for a single SPDC crystal}
To calculate the generation rate of either the signal or idler photons in one crystal lamella, we use \(R_j= \langle b_j^\dag (t,\vec r) b_j(t,\vec r) \rangle\). Fourier transforming from \(t,x,y\rightarrow \omega, k_x, k_y\), and using the solution in Eq. \eqref{Crystal operators solution} yields after some algebra: 
\begin{equation}
\begin{split}
R_{si}= &\langle b_{si}^\dag (t,\vec r) b_{si}(t,\vec r) \rangle=\frac{1}{(2\pi)^3}\int_{\omega}\int_{q} |B|^2 dw, dq \\&+ \frac{1}{(2\pi)^3}\int_{\omega}\int_{q}\int_{0}^{L} |F(s)|^2 dw, dq, ds,
\end{split}
\end{equation}
\begin{equation}
\begin{split}
R_{id}= &\langle b_{id}^\dag (t,\vec r) b_{id}(t,\vec r) \rangle=\frac{1}{(2\pi)^3}\int_{\omega}\int_{q} |C|^2 dw, dq \\&+ \frac{1}{(2\pi)^3}\int_{\omega}\int_{q}\int_{0}^{L} |G(s)|^2 dw, dq, ds.
\end{split}
\end{equation}

Assuming two ideal detectors, each with an area A larger that of the pumping beam and given that one detector clicks at time \(t_1\) and position \(\mathbf{r_1}\), the probability of the second detector clicking at time \(t_2=t_1+\tau\) and position  \(\mathbf{r_2}=\mathbf{r_1}+\mu\) is given by the Glauber correlation function \(\mathcal{G}^{(2)}\):
\begin{equation}
\mathcal{G}^{(2)}(\tau,\mu)=\langle b_{si}^\dag (t_2,\mathbf{r_2}) b_{id}^\dag(t_1,\mathbf{r_1})b_{id} (t_1,\mathbf{r_1}) b_{si}(t_2,\mathbf{r_2}) \rangle.
\end{equation}
Fourier transforming, and using Eq. \eqref{Crystal operators solution} and Wick's theorem, we obtain after some algebra the temporal correlation:
\begin{equation}
\mathcal{G}^{(2)}(\tau)=\frac{1}{(2\pi)^4} \int_{q}\bigg| \int_{\omega}\phi(\omega,q)e^{i\omega\tau }d\omega\bigg|^2 dq+R_{si}R_{id},
\end{equation}
where \(\phi(\omega,q)=B^*D+\int_0^LF^*(s)H(s)ds\). The term  \(R_{si}R_{id}\) represents the accidental coincidence counts resulting from signal or idler photons. For a short coincidence time window, this term is small relative to the first term and can be safely neglected. The coincidence count rate per unit area [Photons/(sec area)] is then: 

\begin{equation}
\label{Coincidence count rate}
R_{coincidence}=\int \mathcal{G}^{(2)}(\tau)d\tau=\frac{1}{(2\pi)^3}\bigg|\int_{q}\int_{\omega}\phi(\omega,q)d\omega,dq\bigg|^2.
\end{equation}

The coincidence count rate is obtained by multiplying this expression by the pump flux and expressing q in terms of the propagation angles. The integration limits are set according to the angular range of the detectors. 

\subsection{Model for the SU(1,1) interferometer}
Since SU(1,1) interferometers consist of two SPDC crystals, we can now apply the above results to calculate the rates for our SU(1,1) interferometer. Assuming phase membrane with same absorption coefficient and refraction index as the crystals, The matrices representing a lossy phase membrane of thickness d are: 

\begin{equation}
\begin{split}
& M_{m}=\begin{bmatrix}
{\alpha_{si}+i\Delta k_z /2} & 0 \\
0 & {\alpha_{id}-i\Delta k_z /2}
\end{bmatrix}   
\\ &   exp(-M_{m}d)=\begin{bmatrix}
A_m & 0 \\
0 & D_m
\end{bmatrix}
\end{split}
\end{equation}
Adding the Langevin term and following the same procedure as before, the operators after one crystal and the membrane are given by: 
\begin{equation}
\label{Membrane operators solution}
\begin{split}
& {b_{si}}^{c\&m}= A_m(d){b_{si}(L)}+\int_{0}^{d} E_m(s)\sqrt{\alpha_{si}}f_{si}(s)\,ds
\\& {b_{id}^\dag}^{c\&m}=D_m(d){b_{id}^\dag(L)}+\int_{0}^{d} H_m(s)\sqrt{\alpha_{id}}f_{id}^\dag(s)\,ds
\end{split}
\end{equation}
where
\begin{equation}
exp(-M_{m}[s-d])=\begin{bmatrix}
E_m & 0 \\
0 & H_m
\end{bmatrix}
\end{equation}
To obtain the final output after the second identical crystal, we apply Eq. \eqref{Crystal operators solution} once again, using the output from Eq. \eqref{Membrane operators solution} as the input: 
\begin{equation}
\label{final operators solution}
\begin{split}
& {b_{si}}^{SU(1,1)}= A{b_{si}}^{c\&m}+B{b_{id}^\dag}^{c\&m}
\\& \qquad +\int_{0}^{L} E(s)\sqrt{\alpha_{si}}f_{si}(s)\,ds+\int_{0}^{L} F(s)\sqrt{\alpha_{id}}f_{id}^\dag(s)\,ds
\\& {b_{id}^\dag}^{SU(1,1)}= C{b_{si}}^{c\&m}+D{b_{id}^\dag}^{c\&m}
\\& \qquad +\int_{0}^{L} G(s)\sqrt{\alpha_{si}}f_{si}(s)\,ds+\int_{0}^{L} H(s)\sqrt{\alpha_{id}}f_{id}^\dag(s)\,ds
\end{split},
\end{equation}
Similarly to the approach used for a single crystal, this yields the output rates:
\begin{equation}
\begin{split}
R_{si}^{SU(1,1)}= &\langle {b_{si}^\dag}^{SU(1,1)}{b_{si}}^{SU(1,1)} \rangle=\frac{1}{(2\pi)^3}\int_{\omega}\int_{q} {\mathcal{A}}_{si} dw, dq \\&+ \frac{1}{(2\pi)^3}\int_{\omega}\int_{q}\int_{0}^{L} {\mathcal{B}}_{si}(s) dw, dq, ds
\\&+ \frac{1}{(2\pi)^3}\int_{\omega}\int_{q}\int_{0}^{d} {\mathcal{C}}_{si}(s) dw, dq, ds,
\end{split}
\end{equation}
\begin{equation}
\begin{split}
R_{id}^{SU(1,1)}= &\langle {b_{id}^\dag}^{SU(1,1)}{b_{id}}^{SU(1,1)} \rangle=\frac{1}{(2\pi)^3}\int_{\omega}\int_{q} {\mathcal{A}}_{id} dw, dq \\&+ \frac{1}{(2\pi)^3}\int_{\omega}\int_{q}\int_{0}^{L} {\mathcal{B}}_{id}(s) dw, dq, ds
\\&+ \frac{1}{(2\pi)^3}\int_{\omega}\int_{q}\int_{0}^{d} {\mathcal{C}}_{id}(s) dw, dq, ds,
\end{split}
\end{equation}
where: 
\begin{equation}
\begin{split}
&{\mathcal{A}}_{si}=|(AA_m+D_mD)B|^2
\\&{\mathcal{B}}_{si}=|AA_mF+BDH|^2+|F|^2
\\&{\mathcal{C}}_{si}=|BH_m|^2
\\&{\mathcal{A}}_{id}=|CA_mA+DD_mC|^2
\\&{\mathcal{B}}_{id}=|CA_mE+DD_mG|^2+|G|^2
\\&{\mathcal{C}}_{id}=|CE_m|^2
\end{split}.
\end{equation}
Finally, following Eq. \eqref{Coincidence count rate}, the coincidence rate measured after the SU(1,1) interferometer will be:
\begin{equation}
\label{Coincidence SU11 count rate}
R_{coincidence}^{SU(1,1)}=\frac{1}{(2\pi)^3}\bigg|\int_{q}\int_{\omega}\phi^{SU(1,1)}d\omega, dq\bigg|^2.
\end{equation}
where
\begin{equation}
\label{phi su11}
\begin{split}
&\phi^{SU(1,1)}=\mathcal{A}+\int_{0}^{L} \mathcal{B}(s) ds +\int_{0}^{d} \mathcal{C}(s) ds
\\&\mathcal{A}=(A^*A_mB^*+B^*D_mD^*)(CA_mB+DD_mD)
\\&\mathcal{B}=(A^*A_mF^*+B^*D_mH^*)(CA_mF+DD_mH)\\&+F^*H^*
\\&\mathcal{C}=BH_m^*DH_m
\end{split}
\end{equation}

To compare with our experimental results, we use Wolfram Mathematica to numerically integrate Eq. \eqref{Coincidence SU11 count rate} for each membrane thickness d. The red line in Fig. 4 of the main text is the calculated coincidence count rate for membrane thicknesses ranging from 0 to 30 microns.

\section{properties of the interferometer and the phase object}
Figure S.2 shows our interferometer and a typical phase object. Panel (a) is a photograph of the interferometer, with details of its fabrication described in \cite{Deutsch1985AMethod,Deutsch1985ElectronicSilicon}. It is a monolithic device cut from a perfect silicon single crystal, with lamellas \(\sim0.25\)  mm thick, 12 mm high, 25 mm wide and spaced 5mm apart. The (1,1,1) reflection plane is normal to the lamellas, as shown in Fig. S.2(a). The silicon membranes (Fig. S.2(c)) were obtained from Norcada Inc in three thicknesses as described in Table S.1. By stacking these membranes in various combinations, phase objects of thicknesses ranging from 2 to 28 microns were obtained. This approach allowed precise control over the beam's phase modulation. 
 
 \begin{figure}
    \centering \includegraphics[width=1\linewidth]{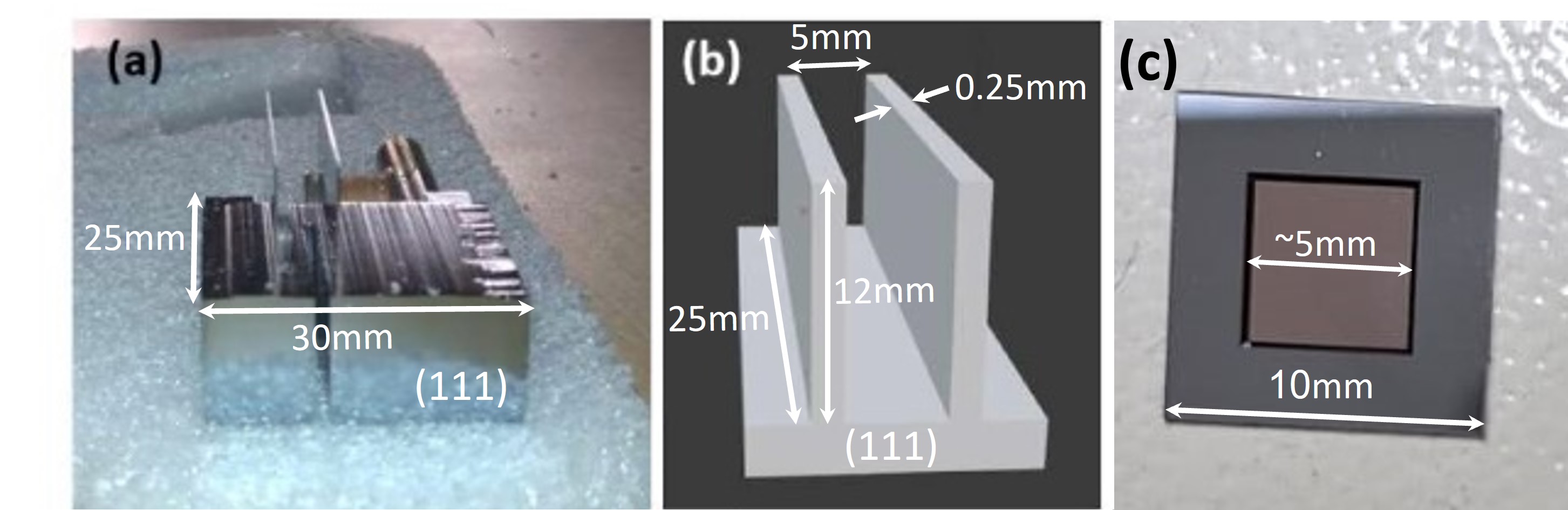}
    \caption {(a,b) The interferometer, and (c) a typical membrane phase object (c)}
    \label{fig:Interferometer_and_membrane}
\end{figure}

\begin{center}
 \begin{tabular}{ | m{1.5cm} | m{0.8cm}| m{0.8cm} |m{1.5cm} |m{1.5cm} | } 
 \hline
 Model & Frame size (mm) & Frame thickness \((\mu m)\) & Membrane Thickness \((\mu m)\) & Membrane dimensions (mm) \\  
 \hline
 SM10480N & \(10{\times}10\) \((\pm0.2)\) & \(300\) \((\pm10)\) & \(2\) \((\pm0.5)\) & \(4.8{\times}4.8\) \((\pm0.05)\) \\ 
 \hline
 SM10500P & \(10{\times}10\) \((\pm0.2)\) & \(300\) \((\pm10)\) & \(10\) \((\pm0.5)\) & \(5{\times}5\) \((\pm0.05)\) \\
 \hline
SM10500Q & \(10{\times}10\) \((\pm0.2)\) & \(300\) \((\pm10)\) & \(20\) \((\pm0.5)\) & \(5{\times}5\) \((\pm0.05)\) \\
 \hline
 \end{tabular}
  \captionof{table}{Dimensions of the membranes used, in combination, as a phase shift object}
\end{center}

\bibliography{references_supplimental}
\bibliographystyle{naturemag}